\documentclass[12pt]{article}

\usepackage[totalheight = 23cm, totalwidth = 17cm]{geometry}
\usepackage{amssymb,amsmath,amsfonts,amsbsy,graphicx,bm}

\usepackage{color,cancel,ulem}

\newcommand{\dis}[1]{\begin{equation}\begin{split}#1\end{split}\end{equation}}

\begin{document}

\begin{titlepage}

\begin{center}

{\LARGE \bf 
 Black hole production, eternal inflation, and information  in quasi-de Sitter space
}

\vskip 1.0cm

{\large
Min-Seok Seo$^{a}$ 
}

\vskip 0.5cm

{\it
$^{a}$Department of Physics Education, Korea National University of Education,
\\ 
Cheongju 28173, Republic of Korea
}

\vskip 1.2cm

\end{center}

\begin{abstract}

When the slow-roll parameter $\epsilon_H$ is smaller than $H^2/M_{\rm Pl}^2$, the quantum fluctuations of the inflaton after the horizon crossing are large enough to realize eternal inflation.
Whereas they do not generate a sufficient amount of density fluctuation of the inflaton to produce the black hole in quasi-de Sitter space, they can also generate the sizeable density fluctuation of the radiation when the number of degrees of freedom increases rapidly in time, as predicted by the distance conjecture.
We argue that the condition that the density fluctuation of the radiation is not large enough to produce the black hole until the end of inflation is equivalent to the no eternal inflation condition.
When the  radiation emitted by the horizon does not produce the black hole, even if the number of degrees of freedom increases in time, the information paradox does not arise for  $\epsilon_H$ larger than $10^{-7} (H^2/ M_{\rm Pl}^2)$ and  time scale shorter than $10^4 (M_{\rm Pl}/H^2)$.
 Regardless of the presence of the information paradox, a static observer cannot retrieve a sufficient amount of information, which is consistent with the complementarity.

\end{abstract}

\end{titlepage}

\newpage

\section{Introduction}

 Inflation, an accelerated expansion of  the early universe, not just accounts for very special initial conditions of the hot big bang \cite{Guth:1980zm, Linde:1981mu, Albrecht:1982wi} but also provides a mechanism to create large scale inhomogeneities from the quantum fluctuations \cite{Mukhanov:1981xt, Mukhanov:1990me}.
 Since various inflationary models make specific predictions for the behavior of quantum fluctuations, testing the models by comparison with observation is expected to be a window into quantum gravity.
 In particular, the strong quantum gravity effects are predicted in models where the horizon radius $H^{-1}$ varies very slowly as characterized by extremely small value of the slow-roll parameter $\epsilon_H \equiv -\dot{H}/H^2$.
 For instance, if $\epsilon_H$ is smaller than $H^2/M_{\rm Pl}^2$, \footnote{Throughout this article, Newton's constant $G$ is identified with $M_{\rm Pl}^{-2}$.  }
 the classical motion of  the inflaton decreasing the vacuum energy density is overwhelmed by the quantum fluctuations in a non-negligible number of causal patches.
 Then the whole universe keeps inflating, realizing eternal inflation   \cite{Steinhardt:1982kg, Vilenkin:1983xq, Linde:1986fc, Linde:1986fd, Goncharov:1987ir} (for a review, see, e.g., \cite{Guth:2007ng}).

 For the  tiny value of $\epsilon_H$  the spacetime geometry  is maintained close to de Sitter (dS) space for a long period of inflation, implying that dS space is (meta)stable against quantum gravity effects.
However, the difficulty of dS model building in string theory has raised doubt that the metastable dS vacuum is in fact the result of fine-tuning \cite{Kachru:2003aw, Balasubramanian:2005zx} or  even worse, not allowed by quantum gravity \cite{Obied:2018sgi, Andriot:2018mav, Garg:2018reu, Ooguri:2018wrx}. 
 This has motivated the intense study on the physical meaning of the condition for (no) eternal inflation in light of  quantum gravity \cite{Arkani-Hamed:2007ryv, Arkani-Hamed:2008mpk, Olum:2012bn, Matsui:2018bsy, Kinney:2018kew, Brahma:2019iyy, Rudelius:2019cfh, Bedroya:2020rac, Seo:2021bpb, Rudelius:2021azq}.
 
 Since we do not completely understand the UV completion of quantum gravity, much of the discussions on the dS instability and the lower bound on $\epsilon_H$ rely on the `swampland conjectures', the conjectured quantum gravity constraints on the low energy effective field theory (EFT) (for reviews, see, e.g., \cite{Brennan:2017rbf, Palti:2019pca, vanBeest:2021lhn, Grana:2021zvf}).
 A number of conjectures claim that the universe as we observe it is not the result of fine-tuning, but must be the unavoidable consequence of the fundamental principle  of quantum gravity.
  In particular,  the stability of the EFT  against the large quantum gravity effects has often been conjectured.
  Conjectures in this direction include the no black hole excitation from the vacuum \cite{Cohen:1998zx}, the weakness of gravity \cite{Arkani-Hamed:2006emk}, and the short period of inflation preventing the horizon crossing of trans-Planckian modes \cite{Bedroya:2019snp}.
  If they are true, eternal inflation requiring the dominance of the quantum fluctuations over the semiclassical behavior controlled by the EFT may be forbidden by the as-yet-unknown quantum gravity effects.
  
 As the large quantum fluctuations not only give rise to eternal inflation but also excite the black hole state from the vacuum, we expect the close relationship between   two phenomena.
 Regarding this issue, a remarkable conjecture was proposed by  Cohen, Kaplan, and Nelson (CKN) stating that the UV cutoff of the EFT defined in the finite region must be low enough not to produce the black hole  \cite{Cohen:1998zx}, which can be applied to the subhorizon region in (quasi-)dS space \cite{Banks:2019arz,  Seo:2021bpb, Castellano:2021mmx}.
 Whereas the equivalence between the no eternal inflation and the no black hole production conditions was suggested in \cite{Seo:2021bpb}, it was drawn by  combining the dS CKN bound with the distance conjecture, another swampland conjecture claiming the breakdown of the EFT by the rapid increase of the low energy degrees of freedom along the moduli trajectory \cite{Ooguri:2006in}. 
 Indeed, the distance conjecture was used to argue that the stable dS spacetime belongs to the swampland \cite{Ooguri:2018wrx}.
 More concretely, as the number of low energy degrees of freedom rapidly increases in time, the entropy of matter produced inside the horizon eventually exceeds the covariant entropy bound given by the area of the horizon in Planck unit.
 In order that the covariant entropy bound is not violated, the spacetime geometry would be deformed from dS space. 
 Since this `dS swampland conjecture' predicts that $\epsilon_H$ soon becomes ${\cal O}(1)$ even if the spacetime geometry is initially close to dS space,  eternal inflation is expected to be  forbidden as pointed out in, e.g., \cite{Matsui:2018bsy, Kinney:2018kew, Brahma:2019iyy, Rudelius:2019cfh}.
 In this regard, the distance conjecture may be closely connected to the black hole production condition in dS space which arises from the large quantum fluctuation effects like eternal inflation.

To see this more concretely, the first part of this article is devoted to revisiting the issue of the  no black hole production condition from different point of view and comparing it with the no eternal inflation condition.
For this purpose, instead of considering the UV/IR mixing as done in \cite{Cohen:1998zx}, we investigate the amount of the density fluctuations needed for the black hole production.
Regarding this, it is well known that  the quantum fluctuations of the inflaton alone do not generate a sufficient amount of the density fluctuations for the black hole production.
 However, they can also induce the density fluctuations of the radiation, the number of degrees of freedom of which increases in time as predicted by the distance conjecture.
 From this, we find the condition that the black hole is not produced even if the density fluctuations of the radiation are accumulated to the end of  inflation and  argue that it is equivalent to the no eternal inflation condition, $\epsilon_H > H^2/M_{\rm Pl}^2$.

 The radiation we are considering originates from thermal excitations emitted by the horizon.
  A static observer in  (quasi-)dS space will find that the subhorizon region is filled with  thermal excitations  \cite{Gibbons:1977mu}.
  Unlike the evaporating black hole, however, it does not necessarily mean that the entropy of the thermal excitations monotonically increases in time and exceeds the geometric entropy, leading to the information paradox \cite{Hawking:1976ra}.
  In perfect dS space,  dS isometries impose the thermal equilibrium between the thermal excitations and the geometry.
  Since both the thermal and the geometric entropies do not evolve in time,  the information paradox does not arise.
  In the quasi-dS background for the slow-roll inflation where some of dS isometries are spontaneously broken, entropies evolve in time but as we will see, the   information paradox  arises  only when the number of degrees of freedom rapidly  increases in time, as predicted by the distance conjecture (see also \cite{Seo:2022ezk}).
  
  If we assume that quantum gravity forbids the production of the black hole from the quantum fluctuations, the thermal excitations do not contain the black hole but consist only of  the radiation.
  In the second part of this article, we  find that in this case, even if the number of degrees of freedom increases in time, the information paradox does not arise for  $\epsilon_H$ larger than $10^{-7} (H^2/ M_{\rm Pl}^2)$ and  time scale shorter than $10^4 (M_{\rm Pl}/H^2)$.
  It is remarkable that this bound on $\epsilon_H$ has the same parametric dependence as the no eternal inflation condition despite the much suppressed  numerical factor.
  We then close our discussion with the remark that regardless of the appearance of the information paradox, the static observer is free of the potential problem that complementarity \cite{Susskind:1993if, Susskind:1993mu, Hayden:2007cs, Sekino:2008he}  can be violated in the slow-roll background (see also \cite{Aalsma:2020aib, Bedroya:2020rmd} for relevant discussions).

\section{No black hole production condition }

\subsection{Review on quantum fluctuations during inflation}

 We begin the discussion with a review on the behavior of the quantum fluctuations of the inflaton $\phi$ during  inflation. 
 The spacetime geometry during  inflation is well described by   quasi-dS space, the metric of which in the  flat coordinates is written as
 \dis{ds^2=-dt^2+a(t)^2  (dr^2+r^2d\Omega_2^2).\label{eq:dSflat}}
 The classical trajectory of the inflaton $\phi(t)$ depends only on $t$ and $H =\dot{a}/a$ becomes a constant in perfect dS limit.
  Since the dS isometry associated with the time translation and the spatial rescaling is spontaneously broken, the quantum fluctuation of the trace part of the metric  is combined with that of the inflaton to form a physical and gauge invariant  fluctuation  given by \cite{Mukhanov:1985rz, Sasaki:1986hm}  
 \dis{\varphi(t, {\mathbf x}) = \delta \phi (t, {\mathbf x}) -\frac{{\dot \phi}(t)}{H}\frac{{\delta a}(t, {\mathbf x})}{a(t)}.}
 We note that $\delta a/a$, or $\delta N_e = H\delta t$ is interpreted as the quantum fluctuation of the metric in the direction of the time translation, the spontaneously broken dS isometry.
When the nonlinear interactions are negligibly small and $\epsilon_H \ll 1$, the mode expansion
 \dis{\varphi(t, {\mathbf x})=\int \frac{d^3 k}{(2\pi)^3}\frac{e^{i {\mathbf k}\cdot {\mathbf x}}}{\sqrt{2k}a}\Big[e^{-i k\tau}\Big(1-\frac{i}{k\tau}\Big)a_{\mathbf k}+e^{i k\tau}\Big(1+\frac{i}{k\tau}\Big)a^\dagger_{-{\mathbf k}}\Big],
\label{eq:varphi} } 
  where $\tau=-(aH)^{-1}$ is the conformal time, is a good description, from which the two-point correlator is given by
 \dis{\langle \varphi_{\mathbf k} \varphi_{{\mathbf k}'}\rangle = \frac{H^2}{2 k^3}\Big[1+\frac{k^2}{(a H)^2}\Big](2\pi)^3\delta^3(\mathbf{k}+\mathbf{k}').}
 As the universe expands, the wavelength of the mode  $\varphi_{\mathbf k}$ is stretched such that after $t = H^{-1}\log(k/H)$ at which $k = aH$ is satisfied (horizon crossing), the amplitude of $\varphi_{\mathbf k}$ is frozen.
 Then $\varphi_{\mathbf k}$ behaves like the fluctuation of the  classical trajectory $\phi(t)$ as the quantum interference effects become suppressed \cite{Burgess:2006jn, Burgess:2014eoa, Nelson:2016kjm, Shandera:2017qkg, Gong:2019yyz}.
 This generates the  accumulated uncertainty of $\phi(t)$  given by \cite{Vilenkin:1982wt, Linde:1982uu, Starobinsky:1982ee}
 \dis{\langle \phi(t)^2\rangle \Big|_{t_i}^{t_f} = \int \frac{d^3 k}{(2\pi)^3}\frac{d^3 k'}{(2\pi)^3}\langle \varphi_{\mathbf k}\varphi_{{\mathbf k}'}\rangle 
 =\Big(\frac{H}{2\pi}\Big)^2 \log\Big(\frac{k_f}{k_i}\Big),}
 during the time interval $t_f-t_i$ ($k_i = a(t_i)H$ and $k_f = a(t_f)H$), where the contribution of $k/(aH)$ which is much smaller than $1$ after the horizon crossing to the integration is suppressed.
 Since $\log(k_f/k_i)$ is interpreted as the number of $e$-folds $\Delta N_e \simeq H(t_f-t_i)$, the uncertainty of $\phi(t)$ generated by the frozen quantum fluctuations per unit $e$-fold is given by $H/(2\pi)$ which will be denoted by $\Delta$.
 
 In the absence of such uncertainty, the displacement of the inflaton $\phi(\Delta t)$ during $\Delta t$ is fixed to be $-\dot{\phi}\Delta t=-(\dot{\phi}/H)\Delta N_e$.
 If the frozen quantum fluctuations accumulated during the same time interval are large enough to compensate  $(\dot{\phi}/H)\Delta N_e$ in at least one of causal patches,  the vacuum energy density in this patch does not decrease in time as the inflaton does not roll down.
 Then the patch keeps inflating, thus the whole universe also expands, realizing  eternal inflation. 
  Assuming that the fluctuation of $\phi(\Delta t)$ generated in this way obeys the Gaussian distribution, the condition for  eternal inflation to take place after $\Delta N_e$ is given by
 \dis{P\big(\phi >\frac{\dot \phi}{H}\Delta N_e\big) &= \int_{({\dot \phi}/H)\Delta N_e}^\infty d \phi \frac{1}{\sqrt{2\pi}\Delta (\Delta N_e)^{1/2}}e^{-\frac{\phi^2}{2 \Delta^2 (\Delta N_e)}}=\frac12{\rm erfc}\Big(\frac{({\dot \phi}/H)\Delta N_e}{\Delta (\Delta N_e)^{1/2}}\Big)
 \\
 & > e^{-3 (\Delta N_e) } ,}
since a single causal patch becomes $e^{3 (\Delta N_e)}$ patches after  $\Delta N_e$.
From 
\dis{\epsilon_H=\frac{4\pi \dot{\phi}^2}{M_{\rm Pl}^2 H^2}\quad\quad &{\rm and}
\\
{\rm erfc}(x)\simeq \frac{e^{-x^2}}{x\sqrt{\pi}}\Big(1-\frac{1}{2 x^2}+\cdots\Big)\quad\quad &{\rm for}~|x|\gg 1,}
we find that this condition is roughly estimated as  
\footnote{See also \cite{Seo:2020ger} for the possibility that the   bound on $\epsilon_H$ allowing eternal inflation is given by the integer multiple of $H^2/M_{\rm Pl}^2$ when the universe is in the quantum mechanically excited state.}
\dis{\epsilon_H \lesssim \frac{3}{\pi}\frac{H^2}{M_{\rm Pl}^2}.\label{eq:etcond}}

Meanwhile, the frozen quantum fluctuations are also related to the density fluctuations.
This can be seen  by considering  the Einstein equation $\delta G_{\mu\nu}=-8\pi M_{\rm Pl}^{-2} \delta T_{\mu\nu}$ giving (see, e.g., Sec. 8.3 of \cite{Mukhanov:2005sc})
\dis{\frac{\delta\rho}{\rho}=\frac23 \epsilon_H\Big[\frac{d}{d N_e}\Big(\frac{H}{\dot \phi}\varphi\Big)-3\frac{H}{\dot \phi}\varphi\Big].\label{eq:drhorho}}
The proportionality to $\epsilon_H$ can be easily understood by noting that the fluctuation of the energy density $\rho = [3/(8\pi)]M_{\rm Pl}^2 H^2$ under the fluctuation of time $\delta N_e \sim (H/{\dot\phi})\varphi$ (also known as the curvature perturbation) is given by $\delta \rho = [3/(4\pi)]M_{\rm Pl}^2 {\dot H}\delta N_e$, or equivalently, $-2\epsilon_H \rho\delta N_e$, which exactly coincides with the second term.
On the other hand, using  $d\tau/dt = a^{-1}$, we find
\dis{\frac{d\varphi}{d N_e}=-i\int \frac{d^3 k}{(2\pi)^3}\frac{e^{i {\mathbf k}\cdot {\mathbf x}}}{\sqrt{2k}a}\frac{k}{aH}\Big[e^{-i k\tau}a_{\mathbf k}-e^{i k\tau} a^\dagger_{-{\mathbf k}}\Big].}
This is much suppressed for the frozen quantum fluatuations, i.e., the modes after the horizon crossing ($k/(aH) \ll 1$).
Then we find that the first term in \eqref{eq:drhorho} is suppressed as well.
Thus, the frozen quantum fluctuations of $\varphi$ contribute to the density fluctuations  as
\dis{\Big\langle \Big(\frac{\delta\rho}{\rho}\Big)^2\Big\rangle_{\Delta N_e} =4\epsilon_H^2 \frac{H^2}{{\dot\phi}^2}\Big(\frac{H}{2\pi}\Big)^2\Delta N_e = \frac{4}\pi{\epsilon_H}\frac{H^2}{M_{\rm Pl}^2}\Delta N_e.\label{eq:rhofluc}} 

We close this section with the comment on the well-known argument connecting the  eternal inflation condition to  $\delta N_e$ (see, e.g., \cite{Arkani-Hamed:2007ryv}).
This is based on the observation that the frozen quantum fluctuations reenter the horizon in the radiation dominated era, in which $\delta \rho/\rho$ is identified with $\delta N_e=\delta a/a$.
Then the primordial black hole can be created if $\delta \rho/\rho $, or equivalently,  $\delta N_e \sim (H/\dot{\phi})\varphi$ produced during the typical time scale $\Delta N_e\sim {\cal O}(1)$ given by $H/(\epsilon_H^{1/2} M_{\rm Pl})$ becomes ${\cal O}(1)$. 
This evidently shows that the condition for the primordial black hole production is nothing more than the eternal inflation condition $\epsilon_H<H^2/M_{\rm Pl}^2$.
Here we would like to point out that $\delta \rho/\rho$ can be identified with $\delta a/a$ because the universe is in the {\it radiation dominated era}, not in the inflationary era.
We can see this from $\delta \rho = [3/(4\pi)]M_{\rm Pl}^2 {\dot H}\delta N_e$ : when the universe is dominated by radiation, the Hubble parameter is given by $H=(2t)^{-1}$, resulting in $\dot{H}=-2 H^2$, from which we obtain $|\delta \rho/\rho|=4 \delta N_e \sim \delta a/a$.
Here $\delta N_e$ comes from the reentered modes which were frozen during the inflationary era.
In contrary, in the inflationary era, $|\delta \rho/\rho|$ becomes $2\epsilon_H \delta N_e$, in which a factor $\epsilon_H$  originates from the slow change of $H$ in time.
This indeed is consistent with our intuition that the large amount of density fluctuation will be soon diluted away by the exponential expansion of the universe.
 
 In our discussion, we will focus on the black hole production in (quasi-)dS space, not in the geometry in the post-inflationary era.
Then the standard argument which considers the radiation dominated era is irrelevant to our discussion.  
 We also note that in order that the black hole can be produced in (quasi-)dS space, we need an additional mechanism which allows the concentration of matter against the dilution through the exponential expansion.
 One way to achieve this might be the rapid increase of the number of degrees of freedom, in which case a large amount of density fluctuation consisting of a large number of degrees of freedom, not just the fluctuation of $\varphi$ can be produced at once.
 This indeed is predicted by the distance conjecture, which will be addressed in later discussion.

\subsection{No black hole production condition from inflaton fluctuations}

To find the condition that the accumulation of the frozen quantum fluctuations during inflation cannot produce the black hole, we consider the simplest case, the uncharged and nonrotating black hole in the (quasi-)dS background described by the Schwarzschild-de Sitter solution,
\dis{&ds^2=-f(r_s) dt_s^2+\frac{1}{f(r_s)} dr_s^2 +r_s^2d\Omega_2^2,
\\
&f(r_s)=1-\frac{2GM}{r_s}-H^2 r_s^2.}
Requiring $f(r_s) \geq 0$ in order to hide the singularity behind the black hole horizon, we find that the  value of the `black hole mass' $M$ is restricted to satisfy $0 \leq 2 GM \leq (2/3^{3/2})H^{-1}$.
Indeed, as  $M$ increases, the black hole horizon $r_1$ also increases  but at the same time, the black hole backreacts on the geometry such that the (cosmological) horizon $r_2$  decreases until it coincides with $r_1$.
Thus, the upper bound on $M$ is saturated when $r_1=r_2$, which corresponds to the Nariai black hole, the largest realistic black hole in the dS background.
This becomes evident by expressing $M$, $r_1$ and  $r_2$ in terms of a single parameter $\theta \in [\pi/2, \pi]$ as
\dis{&GM=-\frac{1}{3^{3/2} H}\cos\theta,\quad
Hr_1= -\frac{2}{\sqrt3}\cos\big(\frac{\pi+\theta}{3}\big),\quad
Hr_2=\frac{2}{\sqrt3}\cos\frac{\theta}{3},\label{eq:SdSBH}}
respectively.
Here $r_1$ ($r_2$) is a monotonically increasing (decreasing) function of $ \theta$.
For $M$ close to $0$, or equivalently, $\theta$  close to $\pi/2$, two horizon radii are approximated as $r_1 \simeq 2GM$ and $r_2 \simeq H^{-1}-GM$, respectively. 
We also find that for the Nariai black hole  ($\theta=\pi$), two horizons coincide, giving $r_1=r_2=(1/\sqrt3)H^{-1}$.

 We note that whereas the black hole mass has an upper bound in the  dS background, even small mass can produce a black hole if we can put the whole mass into the region of the size $r_1$ for the given mass.
 Hence, to see if the frozen quantum fluctuations can produce the black hole in the quasi-dS background, it is reasonable to investigate whether the size of the density fluctuation given by \eqref{eq:rhofluc} exceeds the density of the black hole, rather than considering the mass produced.
 From \eqref{eq:SdSBH}, the density of the black hole is given by
 \dis{\rho_B=\frac{M}{\frac43 \pi r_1^3}=\frac{3}{32\pi}M_{\rm Pl}^2 H^2\frac{\cos\theta}{\cos^3\big(\frac{\pi+\theta}{3}\big)},}
 which is monotonically decreasing over $\pi/2 \leq \theta \leq \pi$.
 As $\theta \to \pi/2$, $\rho_B$ diverges, which means that the small mass can produce a black hole only if it is concentrated in   the  extremely tiny region. 
 On the other hand, whereas the Nariai black hole is the heaviest black hole in the dS background, it has the smallest   density given by  $\rho_{\rm NB} \equiv [3/(4\pi)]M_{\rm Pl}^2 H^2$, or $\rho_{\rm NB}/\rho=2$.
 Then for  the density fluctuation $\delta \rho/\rho$ in quasi-dS space to produce a black hole,  it is required to be larger than $\rho_{\rm NB}/\rho = 2$.

 Meanwhile, whereas  the density fluctuation given by \eqref{eq:rhofluc}   increases in time as the frozen quantum fluctuations of $\varphi$ are accumulated, it is not sufficient to produce the black hole due to the suppression by $\epsilon_H$.
 That is, if the  black hole is not produced until  $\Delta N_e$ has passed, the accumulated density fluctuation $\langle(\delta \rho/\rho)^2\rangle_{\Delta N_e}$  must be smaller than $(\rho_{\rm NB}/\rho)^2$, the minimum value of $(\rho_B/\rho)^2$, which is written as
 \dis{\frac{4}{\pi}\epsilon_H \Big(\frac{H}{M_{\rm Pl}}\Big)^2\Delta N_e < 4.\label{eq:BHbound}}
 The assumption of the above expression is that $\epsilon_H$ is very small so $H$ can be treated as a constant during $\Delta N_e$.
 This means that $\Delta N_e$ we are considering is smaller than $1/\epsilon_H$ after which the value of $H$ considerably deviates from the initial value.
 Indeed, for this reason, we expect that the total number of $e$-folds during inflation is given by ${\cal O}(1/\epsilon_H)$.
 Thus,  if we require that the black hole is not produced during the whole period of inflation, we may replace $\Delta N_e$ in LHS of \eqref{eq:BHbound} by $1/\epsilon_H$ up to constant, which results in a trivial relation 
 \dis{\frac{H^2}{M_{\rm Pl}^2} < {\cal O}(1).}
 This does not give any constraint to the value of $\epsilon_H$.

 \subsection{No black hole production condition from radiation and distance conjecture}
 
 We now consider the thermodynamics  as seen by a static observer, who is surrounded by the horizon emitting the radiation.
  The static observer describes the background using the static coordinates, in terms of which the metric is given by 
 \dis{ds^2=-(1-H^2 r_s^2) dt_s^2 + \frac{1}{(1-H^2 r_s^2)}dr_s^2+r_s^2 d\Omega_2^2.}
   Comparing with the flat coordinates associated with the metric \eqref{eq:dSflat}, two coordinates are related as
  \dis{e^{-Ht_s}=e^{-Ht}\sqrt{1-H^2 r_s^2},\quad\quad r_s=r e^{Ht}.}
 For  the static observer staying at  fixed $r_s$,  say, $r_s=0$,   the derivatives with respect to $t$ and $t_s$ are not distinguished since $dt =dt_s$.
 
   As pointed out by Gibbons and Hawking, a static observer in  dS space finds that the region beyond the horizon behaves like the blackbody with temperature $T_0=H/(2\pi)$ and entropy $S_{\rm dS}=\pi M_{\rm Pl}^2/H^2$ \cite{Gibbons:1977mu}. 
  Then the region inside the horizon is described to be  filled with the Gibbons-Hawking radiation.
  Taking the massless bosonic radiation into account   for simplicity, and noting that the temperature is blueshifted as $T(r_s)=T_0/\sqrt{1-H^2r_s^2}$ depending on $r_s$, the energy of the radiation is given by
  \dis{E_{\rm rad}&=\int dV\rho ={\cal N}\int \frac{d\Omega_2 dr r^2}{\sqrt{1-H^2r^2}} \int \frac{d\Omega_2 dp p^2}{(2\pi)^3} \frac{p}{e^{p/T(r)}-1}
\\
&= \frac{{\cal N}}{2}\Big(\frac{H}{2\pi}\Big)^4\frac{\Gamma(4)\zeta(4)}{\Gamma(\frac32)^2}\int_0^{H^{-1}}\frac{r^2}{(1-H^2r^2)^{5/2}}dr,}
 where ${\cal N}$ is the number of massless degrees of freedom and we omitted the subscript $s$ representing the static coordinates.
  The integration over $r$ diverges due to the blueshift, but we may regularize it by replacing the upper bound of the integration $H^{-1}$ by $H^{-1}-\Lambda_{\rm UV}^{-1}$.
  Taking $\Lambda_{\rm UV}$ to be $M_{\rm Pl}$, we obtain
  \footnote{One may take $\Lambda_{\rm UV}$ to be $M_{\rm Pl}^2/H$ by requiring  the {\it proper distance} between the upper bound of $r_s$ and the horizon to be $M_{\rm Pl}^{-1}$.
  Whereas it is motivated by the stretched horizon of the black hole, the energy density in this case is, up to constant, given by $\rho \sim H M_{\rm Pl}^3$ which is larger than the dS energy density $\sim H^2 M_{\rm Pl}^2$.
  For this reason, we think $M_{\rm Pl}^2/H$ is not an appropriate choice of the cutoff. }  
  \dis{E_{\rm rad}={\cal N}\frac{H}{240\sqrt2 \pi}\Big[\frac13\Big(\frac{M_{\rm Pl}}{H}\Big)^{3/2}-\frac34 \Big(\frac{M_{\rm Pl}}{H}\Big)^{1/2}\Big],}
  where the ${\cal O}((H/M_{\rm Pl})^{1/2})$ terms are suppressed.

 Now suppose ${\cal N}$ increases rapidly in time as predicted by the distance conjecture \cite{Ooguri:2006in} : as the inflaton traverses along the trans-Planckian geodesic distance, infinite towers of states descend from UV, invalidating the EFT.
 In this case, we can consider an ansatz for ${\cal N}$ given by
  \dis{{\cal N}={\cal N}_0 e^{\lambda  \phi(t)/M_{\rm Pl}},\label{eq:Nansatz}}
 where ${\cal N}_0$ is the number of towers in the EFT and $\lambda$ is an ${\cal O}(1)$ constant.
As the frozen quantum fluctuations of $\varphi$ are accumulated,   $\phi(t)$ also fluctuates, which gives rise to the density fluctuation of the radiation,
\dis{\Big(\frac{\delta \rho_{\rm rad}}{\rho_{\rm rad}}\Big)^2 =\Big(\frac{\delta E_{\rm rad}}{E_{\rm rad}}\Big)^2=\lambda^2 \frac{ \langle \phi^2 \rangle_{\Delta N_e}}{M_{\rm Pl}^2}
 = \frac{\lambda^2}{4\pi^2}\frac{H^2}{M_{\rm Pl}^2}\Delta N_e.}
If the accumulation of the density fluctuation above is not large enough to produce the black hole until the end of inflation, this is required to be smaller than $(\rho_{\rm NB}/\rho)^2 = 4$ for $\Delta N_e \sim \epsilon_H^{-1}$.
 Then  $\epsilon_H$ is bounded as
 \dis{\epsilon_H > \frac{\lambda^2}{16\pi^2}\frac{H^2}{M_{\rm Pl}^2},\label{eq:noBHdis}}
 which is equivalent to the no eternal inflation condition.
 This is consistent with the conclusion in \cite{Seo:2021bpb} that the distance conjecture forbids the production of the black hole in dS space.

 \section{Information paradox in quasi-dS space without black hole production }

\subsection{The condition for the absence of information paradox}
   When $S_{\rm dS}$ is interpreted to describe the number of degrees of freedom of the quantum system beyond the cosmological horizon as seen by a static observer, the static observer finds that the Gibbons-Hawking radiation is produced by the quantum fluctuation on the horizon, such as a pair production.
 Then we expect that the Gibbons-Hawking radiation is entangled with the degrees of freedom which recede beyond the horizon, the number of which is counted in $S_{\rm dS}$.
 This implies that the entropy of the Gibbons-Hawking radiation is required to be smaller than $S_{\rm dS}$.
 We may regard such a restriction $S_{\rm rad} < S_{\rm dS}$ as a special case of the covariant entropy bound \cite{Bousso:1999xy}. \
 
 Now suppose $\epsilon_H$ satisfies the bound \eqref{eq:noBHdis} such that the black hole is not produced  in quasi-dS space during the whole period of inflation.
  As we have seen, this is a result of the distance conjecture, which predicts the rapid increase of the number of degrees of freedom ${\cal N}$ as given by \eqref{eq:Nansatz}.
 In the absence of the black hole, the entropy of the Gibbons-Hawking radiation is given by (see also \cite{Parikh:2008iu})
\dis{S_{\rm rad}&=\int dV\frac{\rho+p}{T}=\int \frac{d\Omega_2 dr r^2}{\sqrt{1-H^2r^2}}\frac{{\cal N}}{T(r)}\int \frac{d\Omega_2 dp p^2}{(2\pi)^3}\Big[1+\frac13 \Big]p\frac{1}{e^{p/T(r)}-1}
\\
&=16{\cal N}\Big(\frac{H}{4\pi}\Big)^3\frac{\Gamma(3)\zeta(4)}{\Gamma(\frac32)^2}\int_0^{H^{-1}}\frac{r^2}{(1-H^2r^2)^2}dr.}
As in the case of $E_{\rm rad}$, the integration over $r$ is divergent so we regularize it using the Planck scale cutoff, to obtain
 \dis{S_{\rm rad}=\frac{\cal N}{180}\Big[\frac{M_{\rm Pl}}{H}-\log\Big(\frac{2M_{\rm Pl}}{H}\Big)-\frac12\Big].}
Since the radiation inside the horizon is entangled with the state beyond the horizon, ${\cal N}$ is restricted to satisfy $S_{\rm rad} < S_{\rm dS}$.

In perfect dS space, dS isomerties impose that the energy flux emitted by the horizon is balanced with that absorbed by the horizon.
Hence the radiation is in equilibrium with the background geometry, as  reflected in the time independence of both $S_{\rm dS}$ and $S_{\rm rad}$.
The situation is changed for quasi-dS space, in which some of dS isometries are spontaneously broken by the time evolution of $H$ (for recent discussions, see, e.g., \cite{Aalsma:2019rpt, Gong:2020mbn}). 
As the radiation is no longer in equilibrium with the background, we expect the time evolution of the entropies $S_{\rm rad}$ and $S_{\rm dS}$.
From
\dis{&\frac{dS_{\rm dS}}{dt}=\frac{2\pi \epsilon_H}{H}M_{\rm Pl}^2,
\\
&\frac{dS_{\rm rad}}{dt}=\frac{\cal N}{180}\epsilon_H(M_{\rm Pl}-H)+\frac{d{\cal N}/dt}{180}\Big(\frac{M_{\rm Pl}}{H}-\log\Big(\frac{2M_{\rm Pl}}{H}\Big)-\frac12\Big),}
we find that for $d{\cal N}/dt \geq 0$, both $S_{\rm dS}$ and $S_{\rm rad}$ increase in time.

 If ${\cal N}$ does not evolve in time  ($d{\cal N}/dt=0$),   $S_{\rm dS}$ increases faster than $S_{\rm rad}$, hence $S_{\rm rad}$ never exceeds $S_{\rm dS}$.
  In this case, the information paradox never arises.
  The only way for the information paradox to arise is that ${\cal N}$ increases rapidly in time such that $dS_{\rm rad}/dt$ becomes larger than $dS_{\rm dS}/{dt}$ \cite{Seo:2022ezk}.
 The excess of $S_{\rm rad}$ over $S_{\rm dS}$ in this way indeed is used to argue the instability of dS space \cite{Ooguri:2018wrx}  (for more discussion, see, e.g., \cite{Seo:2019mfk, Seo:2019wsh, Sun:2019obt}). 
 To find $d{\cal N}/dt$ explicitly, we  consider an ansatz motivated by the distance conjecture given by \eqref{eq:Nansatz}.
 Since 
 \dis{  \phi (\Delta t) \simeq {\dot \phi}\Delta t = \sqrt{\frac{\epsilon_H}{4\pi}} M_{\rm Pl}\Delta N_e,\label{eq:phid}}
 we find  $d{\cal N}/d N_e=\lambda [\epsilon_H/(4\pi)]^{1/2}{\cal N}$ is ${\cal O}(\epsilon_H^{1/2})$, which implies that 
 $dS_{\rm rad}/dt$ can be larger than the ${\cal O}(\epsilon_H)$ quantity $dS_{\rm dS}/dt$.
 In this case,  $S_{\rm rad}$ increases to saturate $S_{\rm dS}$ after $\Delta N_e \simeq [\sqrt{4\pi}/(\lambda\sqrt{\epsilon_H})]\log(M_{\rm Pl}/H)$ has passed \cite{Seo:2019wsh, Cai:2019dzj}.

However, even if ${\cal N}$ increases rapidly in this way, the information  paradox cannot arise when $dS_{\rm rad}/dt$ is kept smaller than $dS_{\rm dS}/dt$ by the sizeable value of $\epsilon_H$. 
Comparing the leading terms of  $dS_{\rm dS}/dt$ and $dS_{\rm rad}/dt$, we find that  the no information paradox condition,  $dS_{\rm rad}/dt < dS_{\rm dS}/dt$  is written as 
\dis{\epsilon_H^{1/2} > \frac{1}{720 \pi^{3/2}}{\cal N}_0 e^{\lambda \phi(\Delta t)/M_{\rm Pl}}\frac{H}{M_{\rm Pl}}.\label{eq:entrcom}}
Taking the $\epsilon_H$ dependence of $\phi(\Delta t)$ into account, this bound is equivalent to
\dis{\epsilon_H^{1/2} > -\frac{\sqrt{4\pi}}{\lambda \Delta N_e}W_0\Big(-\frac{\lambda\Delta N_e}{1440 \pi^2}{\cal N}_0\frac{H}{M_{\rm Pl}}\Big),\label{eq:ebound1}}
where $W_0$ is the Lambert W-function, provided the argument of $W_0$ is larger than $-e^{-1}$, or
\dis{\Delta N_e < \frac{1440 \pi^2}{e\lambda{\cal N}_0}\frac{M_{\rm Pl}}{H} \label{eq:ebound2}}
is satisfied.
Since $W_0(x)\simeq x$ for $|x|\ll 1$,  the inequality \eqref{eq:ebound1} is approximated as
\dis{\epsilon_H > \frac{{\cal N}_0^2}{720^2\pi^3}\frac{H^2}{M_{\rm Pl}^2},\label{eq:ebound3}} 
which has the same parametric dependence as the no eternal inflation condition but the numerical factor of ${\cal O}(10^{-7})$ is very tiny for the ${\cal O}(1)$ value of ${\cal N}_0$.
Thus, for $\Delta N_e < {\cal O}(10^4(M_{\rm Pl}/H))$, so far as the black hole is not produced and eternal inflation is forbidden, the information paradox does not arise.
For $\Delta N_e > {\cal O}(10^4(M_{\rm Pl}/H))$, the value of $\epsilon_H$ satisfying \eqref{eq:entrcom} does not exist as \eqref{eq:ebound2} is violated, in which case $dS_{\rm rad}/dt$ is always larger than $dS_{\rm dS}/dt$ as ${\cal N}$ increases rapidly.
Then we expect that   $S_{\rm rad}$ exceeds $S_{\rm dS}$ after $\Delta N_e \simeq [\sqrt{4\pi}/(\lambda\sqrt{\epsilon_H})]\log(M_{\rm Pl}/H)$ has passed in addition, so the information paradox arises.

 We note that $M_{\rm Pl}$ in \eqref{eq:ebound3} in fact is $\Lambda_{\rm UV}$ which regulates the divergences in $E_{\rm rad}$ and $S_{\rm rad}$ 
 (whereas $M_{\rm Pl}$ in \eqref{eq:phid} is irrelevant to $\Lambda_{\rm UV}$,  it does not appear in \eqref{eq:ebound3} since  $\phi(\Delta t)/M_{\rm Pl}$ is written as $[\epsilon_H/(4\pi)]^{1/2}\Delta N_e$).
 While this is a natural choice for the EFT description of gravity, one may take $\Lambda_{\rm UV}$ to be another fundamental scale, say, the string scale.
 This  is lower than $M_{\rm Pl}$, so the bound \eqref{eq:ebound3} is just enhanced, in which case eternal inflation is more easily forbidden.

\subsection{Remarks on information paradox and complementarity}

  When a system in a pure state is divided into two subsystems and an observer can access only one of subsystems, the coarse-graining of the inaccessible subsystem results in the mixed state description of the accessible subsystem. 
  An entanglement or von Neumann  entropy of the accessible subsystem is known to follow the Page curve \cite{Page:1993wv, Page:2013dx}.
  That is, the entanglement entropy increases until the `Page time' at which   two subsystems have the same  number of degrees of freedom, then decreases to zero.
  Meanwhile, it is natural to define information as  the difference between the   equilibrium entropy   and the actual entanglement entropy  since the former and the latter measure the maximal and the actual uncertainties respectively, the difference of which is interpreted as the real certainty  \cite{Page:1993df}.
    In the Page curve, the entanglement entropy before the Page time can be approximated as the equilibrium entropy, thus information is close to $0$.
  This shows that even if information is not lost, an observer should wait until the Page time to get a sufficient amount of information about the inaccessible subsystem.

    For the evaporating black hole, the semiclassical estimation of the radiation entropy follows the equilibrium entropy which is monotonically increasing in time, so we have the information paradox.
   As a resolution,  it has recently been suggested that the radiation entropy follows the Page curve as it will be  purified after the Page time by the contribution from the region inside the black hole horizon  called island  \cite{Penington:2019npb, Almheiri:2019psf, Almheiri:2019hni} (for reviews, see, e.g., \cite{Almheiri:2020cfm, Raju:2020smc}).
  Whereas it provides the description of the black hole and the radiation as seen from far outside the horizon consistent with   unitarity, there also exists the potential issue concerning the no-cloning theorem. 
 To see this, suppose  an observer Alice falls into a black hole carrying the quantum system.
 Since information about the system inside the black hole will be  contained in the radiation after the Page time, it can be collected by another observer Bob who hovers above the black hole horizon.
 If Bob jumps into the black hole carrying  collected information and receives the message from Alice about the system, he has a duplicated state.
 This violates the no-cloning theorem  which reflects the linearity of the unitary evolution.

 A well known resolution to this problem is the complementarity : even though the state can be  cloned, it does not give rise to any problem as long as no observer can find it \cite{Susskind:1993if, Susskind:1993mu, Hayden:2007cs, Sekino:2008he}.
 For Bob to retrieve a sufficient amount of information about Alice, he should wait until at least Page time, which is much longer than the time it takes for the perturbation to the black hole to be scrambled  into the near horizon degrees of freedom.
 Then even though Bob falls into the black hole, he cannot receive the sub-Planckian signal from Alice before reaching  the black hole singularity.
 The trans-Planckian signal is not useful since Bob does not have any tool to analyze it due to his incomplete understanding of quantum gravity.

  We now consider the similar situation in the inflationary cosmology (see also \cite{Nomura:2011dt} for a relevant discussion). 
  When the universe is in the  quasi-dS phase, Alice can recede beyond the horizon carrying the quantum system.
  Suppose  Bob is a static observer.
  If the information paradox arises but is resolved by the island, Bob can have a copy of the state carried by Alice by collecting information contained in the radiation.
    After the end of inflation, the background geometry becomes close to Minkowski space.
  Unlike the  black hole, Minkowski space does not have the singularity and every region is causally connected.
  This means that Bob can receive a message sent by Alice and compare it with the copy he has, which violates the no-cloning theorem.
 The suggestions for the resolution include the following :  the inflation ends   before the scrambling time $H^{-1}\log (S_{\rm dS})$ such that information cannot be contained in the radiation \cite{Bedroya:2019snp, Resol:2019, Bedroya:2020rmd} or Minkowski space again evolves into dS space before Bob receives the message from Alice \cite{Huang:2012eu}.
  
  Indeed,  consideration of the strong subadditivity obeyed by the entanglement entropy shows that  the island does not exist in quasi-dS space \cite{Seo:2022ezk}   as well as perfect dS space \cite{Hartman:2020khs}.
  Since the absence of the island implies that the radiation does not contain information about the state beyond the horizon, Bob cannot  retrieve a sufficient amount of information about the state Alice has.
  In this regard, the absence of the island is consistent with the complementarity but it also means that $S_{\rm rad}$ keeps growing and eventually exceeds $S_{\rm dS}$, which gives rise to the information paradox.
  This can be resolved by demanding that quasi-dS space is strongly deformed to remove the horizon before the information paradox arises, as suggested in  \cite{Ooguri:2018wrx}.
  Another resolution is that, even if the number of degrees of freedom increases rapidly as the distance conjecture predicts, the value of $\epsilon_H$ is bounded such that the increase of $S_{\rm rad}$ is not sufficient to exceed $S_{\rm dS}$. 
  As we have seen, this can be possible when $\epsilon_H$ satisfies both the no black hole production condition and the no information paradox condition  given by  $\epsilon_H >  10^{-7}(H^2/M_{\rm Pl}^2)$ and $\Delta N_e< 10^4 (M_{\rm Pl}/H)$.
  The no information paradox  condition is not satisfied for $\Delta N_e$ longer than $10^4 (M_{\rm Pl}/H)$, after which $S_{\rm rad}$ will saturate  $S_{\rm dS}$.
  In this case, we may impose either the disappearance of the horizon  by the strong deformation or the end of inflation before  $\Delta N_e= 10^4 (M_{\rm Pl}/H)$.
  Presumably, these two may not be independent phenomena.

 The absence of the island in (quasi-)dS space may be connected to the  argument in \cite{Parikh:2008iu} that the radiation in dS space does not contain a sufficient amount of information.
 To be concrete,  even if a black hole is not produced   from the vacuum, since the maximum entropy of the  configuration in a finite region comes from the black hole entropy, we may claim that $S_{\rm rad}$ cannot be larger than the black hole entropy.
  Then the largest entropy of the region inside the  horizon is the entropy of Nariai black hole, the largest black hole in dS space, which is given by  $(\pi/3)(M_{\rm Pl}^2/H^2)$.
  This is similar to $S_{\rm dS}$ in size, but smaller than   $S_{\rm dS}/2$.
 Since the amount of information is less than a single bit even if $S_{\rm rad} = S_{\rm dS}$, Bob cannot  retrieve any one bit of information.

 \section{Conclusions}
\label{sec:conclusion}

 In this article, we obtain the no black hole production condition  by   requiring that even if the number of degrees of freedom increases rapidly in time as the distance conejecture predicts, the accumulated density fluctuation of the radiation is kept smaller than the black hole density during the whole period of inflation $\Delta N_e \sim \epsilon_H^{-1}$.
This   coincides with the no eternal inflation condition  $\epsilon_H > H^2/M_{\rm Pl}^2$, which shows the correlation between eternal inflation and the black hole production under the large amount of frozen quantum fluctuations.

 When the no black hole production condition is satisfied, we expect that the subhorizon region is filled with the radiation  without black hole.
 In this case, the radiation does not give rise to the information paradox if $\epsilon_H > 10^{-7} (H^2/M_{\rm Pl}^2)$ and $\Delta N_e < 10^4 (M_{\rm Pl}/H)$.
 This has the same parametric dependence as the no eternal inflation condition but is much suppressed by the numerical factor.
 Regardless of the presence of the information paradox, a static observer cannot retrieve a sufficient amount of information, which is not contradict to the complementarity.

 We note that whereas we have focused on the density fluctuations to find the no black hole production condition, entropy consideration in  \cite{Chao:1997osu, Bousso:1998na, Bousso:2004tv} suggests that the probability for the black hole production is given by ${\rm exp}[S_{\rm BH}-S_{\rm dS}]$, which is exponentially small.
 This implies that the black hole can be produced in dS space after the exponentially large number of $e$-folds.
 How this modifies our discussion on the no black hole production   condition is the subject of future study.
 Finally, we also note that even though the black hole is not  produced from the frozen quantum fluctuations, it can be formed by the dynamic process like the collapse of the star.
 For the charged black hole in dS space, imposing the absence of the naked singularity during its decay gives the conjectured bound on the charged particle mass \cite{Montero:2019ekk, Montero:2021otb, Lee:2021cor}.
 In this way,  we expect that the evolution of black hole in (quasi-)dS space  may contain some specific aspects of  quantum gravity.

\subsection*{Acknowledgements}

This work is inspired by  an anonymous referee's comment on \cite{Seo:2021bpb}.
'This work was supported by the National Research Foundation of Korea (NRF) grant funded by the Korea government (MSIT) (2021R1A4A5031460).

%

%


\appendix

\renewcommand{\theequation}{\Alph{section}.\arabic{equation}}


\end{document}